\documentclass[conference]{IEEEtran}

\usepackage{cite}
\usepackage{amsmath,amssymb,amsfonts}
\usepackage{graphicx}
\usepackage{booktabs}
\usepackage{multirow}
\usepackage{array}
\usepackage{enumitem}
\usepackage{url}
\usepackage[hidelinks]{hyperref}
\usepackage{balance}

\setlist[itemize]{noitemsep,topsep=1pt,leftmargin=*}
\setlist[enumerate]{noitemsep,topsep=1pt,leftmargin=*}

\newcommand{\model}{Whisper small.en}
\newcommand{\corr}{Corr.}
\newcommand{\rmse}{RMSE}
\newcommand{\epsnorm}{10^{-8}}

\title{Word-Level Modeling with Alignment-Aware Acoustic Fusion for Text-Assisted Intelligibility Prediction in Listeners with Hearing Loss}

\author{
\IEEEauthorblockN{Kazushi Nakazawa}
\IEEEauthorblockA{\textit{Advanced Media, Inc.}\\
Tokyo, Japan}
}

\begin{document}
\bstctlcite{BSTcontrol}
\maketitle

\begin{abstract}
We address text-assisted speech intelligibility prediction for hearing-impaired listeners in CPC3. Although the target is a sentence-level percentage, it is determined by reference-word recognition outcomes. We formulate prediction as reference-conditioned word-level correctness modeling: a frozen Whisper encoder analyzes degraded speech, a teacher-forced decoder conditions on the canonical transcript, and sentence intelligibility is obtained by averaging predicted correctness probabilities over valid reference words. To complement transcript-conditioned decoder states, we add a word-aligned local acoustic branch based on character-level cross-attention alignment and an utterance-level global acoustic branch for calibration. On the official evaluation set, the decoder baseline obtains RMSE 24.92 and correlation 0.795, while joint fusion improves to incorrect-word F1 0.778, MCC 0.626, correlation 0.806, and RMSE 24.39. A similar trend with Whisper medium suggests that the gain comes from prediction granularity and alignment-aware fusion.
\end{abstract}


\section{Introduction}

Objective prediction of speech intelligibility for hearing-impaired listeners is important for hearing-aid evaluation, enhancement-system comparison, and the development of assistive listening technologies. The Clarity Prediction Challenge (CPC) series provides a common benchmark in which systems estimate how many words listeners correctly perceive under realistic hearing-aid and noise conditions \cite{barker22_cpc1,barker24_cpc2,barker2025cpc3}. The third challenge (CPC3) reports intelligibility as a sentence-level percentage, but that percentage is fundamentally an aggregation of word-recognition outcomes.

Many recent systems encode an utterance with a pretrained automatic speech recognition (ASR) model or speech foundation model, pool the hidden representation, and directly regress one sentence-level score \cite{tu22_interspeech,mogridge2024intermediate,cuervo2024sfm,zezario24_interspeech,zhou2025bestpractices}. Such utterance-level regression is simple and effective, yet it creates a granularity mismatch: the training signal is derived from local lexical successes and failures, whereas the model is optimized to predict only one scalar. In the text-assisted protocol considered here, the canonical transcript is available at inference time; therefore, each prediction target has an explicit reference-word anchor. This input condition is different from fully non-intrusive deployment and is made explicit throughout the paper.

\begin{figure}[t]
\centering
\IfFileExists{fig/fig1.pdf}{%
    \includegraphics[width=\columnwidth]{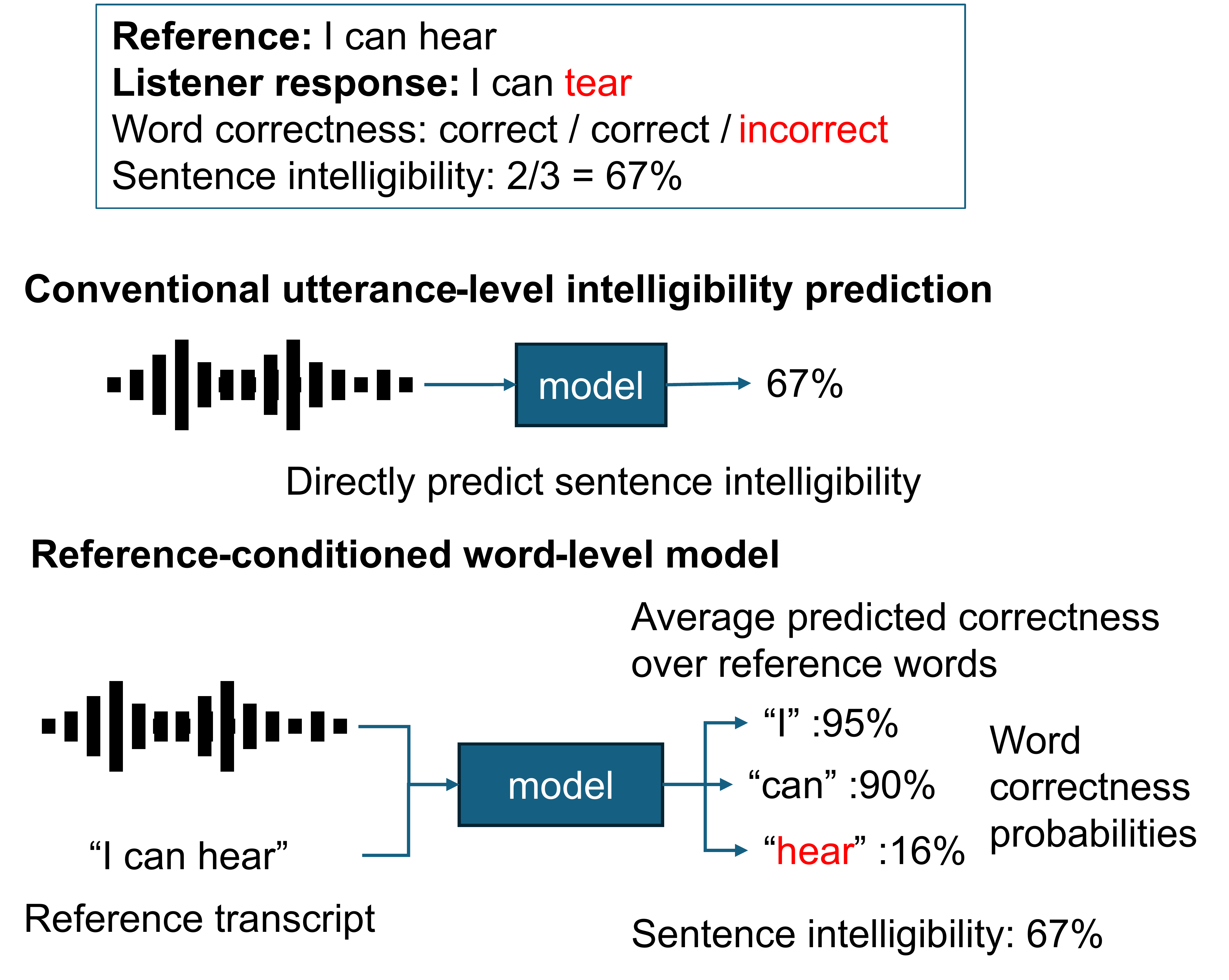}%
}{%
    \fbox{\begin{minipage}{0.92\columnwidth}
    \centering
    \vspace{1mm}
    \textbf{Utterance regression}\\
    speech $\rightarrow$ pooled representation $\rightarrow$ one score\\[1mm]
    \textbf{Reference-conditioned word modeling}\\
    speech + reference text $\rightarrow$ word correctness $\rightarrow$ averaged score
    \vspace{1mm}
    \end{minipage}}
}
\caption{Motivation of reference-conditioned word-level intelligibility prediction. Conventional utterance-level prediction directly regresses the sentence score from speech. In contrast, the proposed formulation predicts correctness probabilities for canonical reference words, such as ``I'', ``can'', and ``hear'', and obtains sentence intelligibility by averaging over valid reference-word positions.}
\label{fig:motivation}
\end{figure}

We address this mismatch by reformulating text-assisted intelligibility prediction as \emph{reference-conditioned word-level correctness modeling}. Rather than using Whisper only as a hypothesis generator, we use it as a frozen audio-text conditional feature extractor \cite{radford2023whisper}. A degraded utterance is encoded by the Whisper encoder, while the decoder is teacher-forced with the canonical transcript. Decoder hidden states are aggregated on reference-word spans, and a classifier predicts whether each reference word was perceived correctly. The sentence-level score is then recovered by averaging correctness probabilities over evaluated words.

This view connects two lines of work. In intelligibility prediction, microscopic and word-level approaches have shown that lexical units are useful targets \cite{best24_interspeech,huckvale25_clarity}. In ASR confidence estimation, token- or word-level reliability estimation benefits from combining decoder-side textual context with local acoustic evidence \cite{swarup19_interspeech,li21_icassp_conf,qiu21_interspeech_conf,naowarat23_interspeech}. Our contribution is a reference-conditioned counterpart of confidence estimation: the model predicts perceptual correctness of canonical reference words for hearing-impaired listeners, not correctness of generated ASR hypotheses.

The proposed model separates three information sources. First, a teacher-forced subword-BPE decoder branch provides semantic reference-word representations. Second, a word-aligned local acoustic branch uses cross-attention maps to extract encoder summaries around each word. Third, an utterance-level global acoustic branch supplies a shared calibration signal for overall acoustic difficulty. To improve local evidence extraction, we use an auxiliary character-level teacher-forced decoder pass and dynamically select alignment-relevant cross-attention heads, motivated by analyses showing that Whisper contains internal alignment heads \cite{yeh2025aligner}. The character-level view is used only for alignment extraction; the main decoder representation remains based on Whisper's standard BPE tokenization.

The main contributions are: (i) a reference-conditioned word-level formulation for CPC3 text-assisted intelligibility prediction; (ii) an alignment-aware local/global acoustic fusion architecture on top of a frozen Whisper backbone; (iii) a controlled analysis showing that character-level dynamic head selection modestly but consistently improves the local branch; and (iv) official evaluation results showing that joint fusion improves both word-level discrimination and sentence-level prediction, with the same qualitative trend preserved for Whisper medium.

\section{Related Work}

Traditional objective intelligibility metrics include intrusive measures such as STOI, HASPI, and HASQI \cite{taal2011stoi,kates2021haspi,kates2022overview}. Recent challenge systems have moved toward non-intrusive or metadata-aware neural prediction, including auditory-model-based approaches \cite{rossbach22_interspeech,mawalim23_clarity}, ASR-derived intermediate representations \cite{tu22_interspeech,mogridge2024intermediate}, speech foundation model features \cite{cuervo2024sfm,zhou2025bestpractices}, and specialized neural architectures such as HASA-Net and MBI-Net \cite{chiang2021hasanet,edozezario22_interspeech}. Related no-reference speech assessment systems, such as NISQA and TorchAudio-Squim, further illustrate the broader shift from hand-crafted intrusive measures toward neural quality and intelligibility estimation \cite{mittag2021nisqa,kumar2023squim}.

A key distinction for the present work is the availability of reference text. Standard non-intrusive assessment is designed for deployment scenarios in which the transcript is unknown, whereas text-assisted evaluation can exploit the canonical utterance as a structured prior. This makes the task closer to reference-word reliability estimation than to unconstrained ASR or generic speech-quality prediction. We therefore treat the transcript not as side information to be pooled at sentence level, but as the coordinate system on which correctness is predicted.

Whisper-based models are now strong baselines for intelligibility prediction because their encoder and decoder states carry robust acoustic and linguistic information \cite{radford2023whisper,zezario24_interspeech}. CPC3 workshop systems further explored multi-domain features, multi-stage training, and heterogeneous Whisper decompositions \cite{buragohain25_clarity,lin25_clarity,zezario25_clarity,jin25_clarity}. Most of these systems, however, remain utterance-level predictors: they pool encoder or decoder states and directly regress the sentence score.

A smaller body of work has examined finer-grained intelligibility modeling. Best \emph{et al.} studied microscopic intelligibility prediction using Whisper transfer learning \cite{best24_interspeech}, and Huckvale proposed a word-level model for CPC3 \cite{huckvale25_clarity}. Our method follows this word-level direction but focuses on a text-assisted encoder-decoder setting in which canonical reference words serve as the prediction coordinates. The local acoustic pathway is also related to alignment extraction. WhisperX uses external alignment components for word timestamps \cite{bain23_interspeech}, while recent work shows that Whisper cross-attention itself can provide internal word-alignment cues, especially with character-based teacher forcing \cite{yeh2025aligner}. We use these cues not for timestamping, but for constructing acoustic summaries that support perceptual correctness prediction.

\section{Method}

\begin{figure*}[t]
\centering
\IfFileExists{fig/fig2.pdf}{%
    \includegraphics[width=\textwidth]{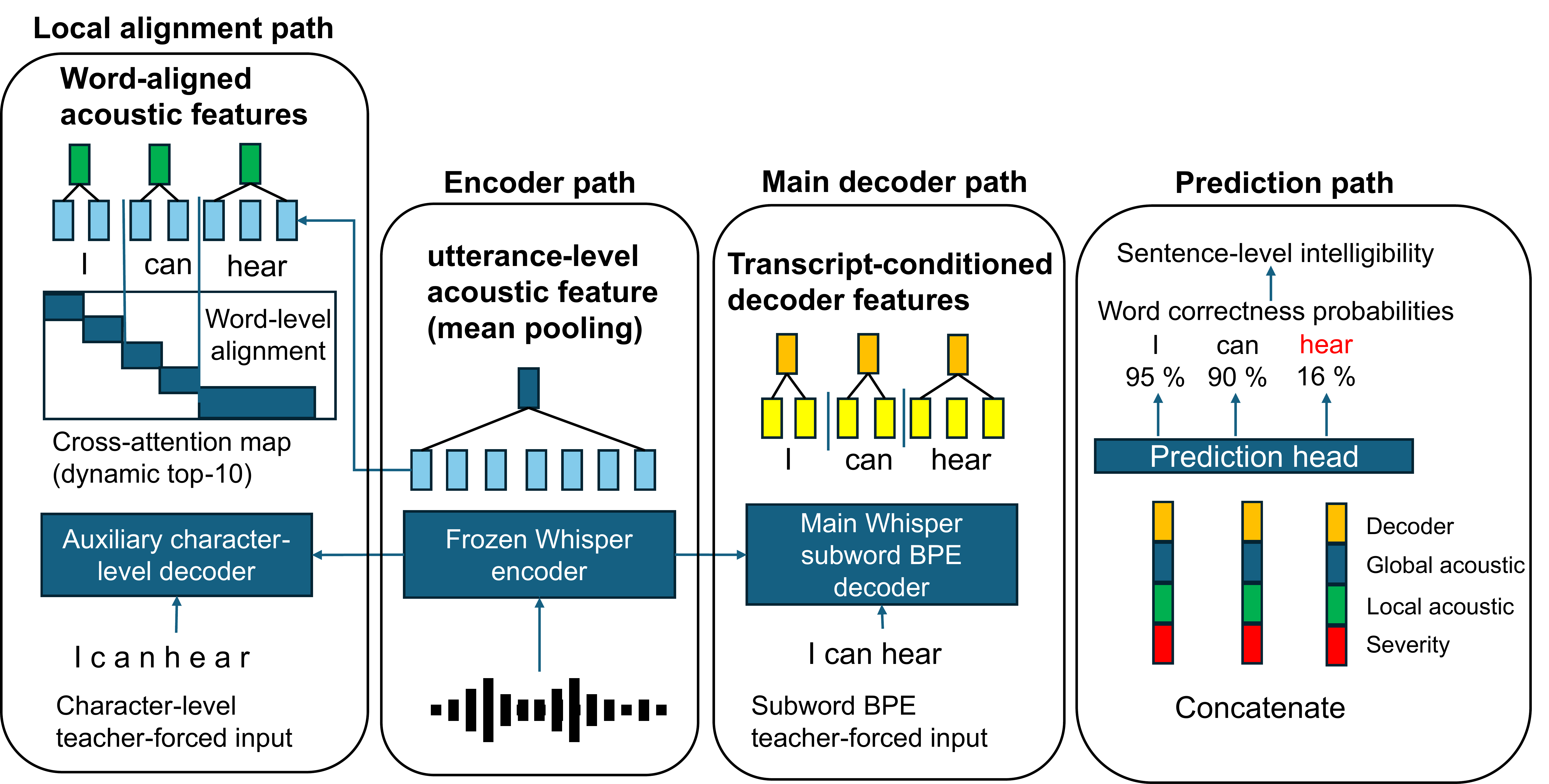}%
}{%
    \fbox{\begin{minipage}{0.94\textwidth}
    \centering
    \vspace{2mm}
    \begin{tabular}{c c c c c}
    Degraded speech & $\rightarrow$ & Frozen Whisper encoder & $\rightarrow$ & encoder states $E$ \\
    Canonical transcript & $\rightarrow$ & teacher-forced BPE decoder & $\rightarrow$ & word states $d_i$ \\
    Canonical transcript & $\rightarrow$ & character-level alignment decoder & $\rightarrow$ & local acoustic states $r_i^{\mathrm{loc}}$ \\
    Encoder states & $\rightarrow$ & masked mean pooling & $\rightarrow$ & global state $g$ \\
    $[d_i,r_i^{\mathrm{loc}},g,\mathrm{severity}]$ & $\rightarrow$ & classifier & $\rightarrow$ & word correctness $p_i$ \\
    \multicolumn{5}{c}{masked mean over valid reference words $\rightarrow$ sentence intelligibility $\hat{y}$}
    \end{tabular}
    \vspace{2mm}
    \end{minipage}}
}
\caption{Overview of alignment-aware multi-granular acoustic fusion. The frozen Whisper encoder provides frame-level acoustic states and an utterance-level global feature. The main Whisper subword-BPE decoder is teacher-forced with the canonical transcript to produce transcript-conditioned decoder features. A separate auxiliary character-level decoder is used only to obtain dynamic top-10 cross-attention maps for word-aligned local acoustic features. The prediction head concatenates decoder, global acoustic, local acoustic, and severity features to estimate word correctness probabilities, whose average gives sentence-level intelligibility.}
\label{fig:framework}
\end{figure*}

Fig.~\ref{fig:framework} summarizes the proposed architecture. The diagram separates the encoder path, the main transcript-conditioned decoder path, the auxiliary local-alignment path, and the final prediction path. All Whisper parameters are frozen. Only projection layers, the severity embedding, and the word-level classifier are trained.

\subsection{Reference-Conditioned Prediction}

Let $x$ be a degraded speech signal, $w_{1:N}$ the canonical transcript represented as reference words, and $s$ the listener severity label. For each evaluated reference word $w_i$, the model predicts a correctness probability $p_i$. The binary target $c_i=1$ indicates that the word was correctly perceived; substitutions and deletions are labeled $0$. Inserted response words do not define new reference targets and are excluded by design. A validity mask $m_i$ indicates which reference words are evaluated. Sentence intelligibility is obtained as
\begin{equation}
\hat{y}=100\cdot \frac{\sum_{i=1}^{N} m_i p_i}{\sum_{i=1}^{N} m_i}.
\label{eq:sent_score}
\end{equation}
Training uses masked binary cross-entropy,
\begin{equation}
\mathcal{L}_{\mathrm{word}}=-\frac{1}{\sum_i m_i}\sum_i m_i\left[c_i\log p_i+(1-c_i)\log(1-p_i)\right].
\label{eq:bce}
\end{equation}

\subsection{Teacher-Forced Decoder Word States}

The Whisper encoder maps $x$ to encoder states $E=(e_1,\ldots,e_L)$, where $e_t\in\mathbb{R}^{D_e}$. The canonical transcript is then teacher-forced through the standard Whisper decoder using its original subword BPE tokenization $t_{1:T}$, producing decoder states $H=(h_1,\ldots,h_T)$. A token-to-word mapping $\mathcal{M}(i)$ assigns decoder tokens to reference words, and the word representation is the mean of decoder states in the corresponding span:
\begin{equation}
 d_i=\frac{1}{|\mathcal{M}(i)|}\sum_{j\in \mathcal{M}(i)} h_j.
\label{eq:word_state}
\end{equation}
Decoder prompt tokens and the start token are excluded from word aggregation. The mapping is constructed by word-aware tokenization with an offset-based fallback when exact token matching is unavailable.

The decoder representation $d_i$ is intentionally transcript-conditioned. It therefore encodes the lexical identity and linguistic context of the reference word, but it can be overly influenced by the supplied text when the degraded audio is ambiguous. The acoustic branches below are designed to counterbalance this effect: the local branch tests whether the aligned speech frames support the reference word, and the global branch summarizes utterance-level difficulty for calibration.

\subsection{Local and Global Acoustic Fusion}

The local branch uses decoder cross-attention as a soft word-to-frame alignment signal. Let $u_{1:U}$ be the auxiliary character-level alignment-token sequence. For decoder layer $\ell$ and head $h$, let $A_{\ell,h}\in\mathbb{R}^{U\times L}$ be the cross-attention matrix. For each utterance, candidate layer-head pairs are scored by the sharpness measure
\begin{equation}
S(A_{\ell,h})=\sum_j\|A_{\ell,h}[j,:]\|_2+\sum_t\|A_{\ell,h}[:,t]\|_2,
\label{eq:sharpness}
\end{equation}
and the top $K=10$ pairs are selected dynamically. The selected maps are averaged, then aggregated over the character positions $\mathcal{C}(i)$ belonging to reference word $w_i$:
\begin{equation}
\tilde{\alpha}_i(t)=\frac{1}{|\mathcal{C}(i)|}\sum_{j\in\mathcal{C}(i)}\frac{1}{K}\sum_{(\ell,h)\in\mathcal{S}}A_{\ell,h}[j,t].
\end{equation}
After temporal normalization, the local acoustic summary is
\begin{equation}
\alpha_i(t)=\frac{\tilde{\alpha}_i(t)}{\sum_{\tau}\tilde{\alpha}_i(\tau)+\epsnorm},\qquad
r_i^{\mathrm{loc}}=\sum_t\alpha_i(t)e_t.
\label{eq:local}
\end{equation}
The normalized profile should be interpreted as a soft context selector. In omission-like cases, there may be no true acoustic realization of the reference word; the mismatch between a strong reference-conditioned decoder state and weak or diffuse local acoustic evidence can itself become informative.

The global branch captures utterance-level difficulty by masked mean pooling the encoder states:
\begin{equation}
g=\mathrm{Pool}(E).
\label{eq:global}
\end{equation}
Decoder, local, and global features are projected to a shared 256-dimensional space and concatenated with a trainable 128-dimensional severity embedding $e_s$. The joint fused vector is
\begin{equation}
z_i=[W_d d_i; W_{\mathrm{loc}} r_i^{\mathrm{loc}}; W_{\mathrm{glob}}g; e_s].
\end{equation}
A lightweight classifier, LayerNorm--Linear--GELU--Dropout--Linear, maps $z_i$ to a logit $\ell_i$, with $p_i=\sigma(\ell_i)$. Decoder-only, local-only, and global-only ablations are obtained by removing the corresponding acoustic features from $z_i$.

\section{Experimental Setup}

We evaluate on CPC3 using the official challenge protocol \cite{barker2025cpc3}. Word-level labels are derived by aligning each listener response to the canonical transcript. Both transcripts are normalized by Unicode NFKC normalization, lowercasing, quote normalization, dash/slash splitting, and punctuation removal except apostrophes. Word-level Levenshtein alignment assigns label $1$ only to exact reference-word matches; substitutions and deletions are labeled $0$.

Input audio is channel-averaged to mono if necessary, resampled to 16 kHz, and truncated to at most 30 seconds before Whisper feature extraction. Unless otherwise stated, the degraded speech waveform is the main input. The oracle-clean alignment condition used in analysis is a diagnostic upper bound and uses clean reference audio only for extracting privileged alignment features.

For model development, we use the official training split and perform five-fold scene-level grouped cross-validation within that split. The organizer-provided development and evaluation sets remain fixed and are not used for selecting checkpoints or hyperparameters. The official evaluation set is used only for final reporting after the model design and checkpoint-selection rule have been fixed. Within each random seed, predictions are averaged across folds. Best checkpoints are selected on the internal validation split using incorrect-word F1, and reported main results are five-seed means with fixed fold assignments.

At sentence level, we report root mean squared error (RMSE) and Pearson correlation. At word level, we report incorrect-word F1, Matthews correlation coefficient (MCC), word accuracy, and sequence exact match. Correctness probabilities are thresholded at 0.5 to obtain binary word predictions. F1 and MCC are then computed with the incorrect class as positive after label inversion.

All main systems use \model{} unless otherwise specified. The task-specific layers are optimized with AdamW, learning rate $10^{-3}$, weight decay $10^{-2}$, linear warmup-decay scheduling, batch size $64$, gradient clipping, dropout $0.1$, mixed precision, and five epochs. Severity conditioning is enabled for all main systems. Trainable parameter counts exclude the frozen Whisper backbone.

\section{Results and Discussion}

\subsection{Main Comparison}

Table~\ref{tab:main} compares the text-conditioned decoder baseline, local fusion, global fusion, and joint fusion under the same frozen backbone, severity conditioning, fold assignments, and optimization settings. The decoder baseline is already strong, reaching RMSE 24.92 and correlation 0.795. This confirms that teacher-forced decoder states provide useful reference-word representations even before explicit acoustic fusion is added. Adding the refined local branch improves incorrect-word F1 from 0.760 to 0.776 and MCC from 0.601 to 0.623, showing that word-aligned acoustic evidence improves lexical failure discrimination. The global branch gives a comparable RMSE reduction with smaller word-level gains, suggesting a calibration role. Joint local/global fusion gives the best overall result, reaching F1 0.778, MCC 0.626, correlation 0.806, and RMSE 24.39.

\begin{table*}[t]
\centering
\caption{Main comparison on the official CPC3 evaluation set using \model{} (five-seed mean $\pm$ standard deviation). Trainable parameters exclude the frozen Whisper backbone. Decoder passes are an architectural inference-cost proxy.}
\label{tab:main}
\scriptsize
\resizebox{\textwidth}{!}{%
\begin{tabular}{lcccccccc}
\toprule
System & Params & Enc. pass & Dec. pass & F1 & MCC & Acc. & \corr & \rmse \\
\midrule
Text-conditioned decoder baseline & 232k & 1 & 1 & 0.760 $\pm$ 0.0005 & 0.601 $\pm$ 0.0005 & 0.807 $\pm$ 0.0002 & 0.795 $\pm$ 0.0003 & 24.92 $\pm$ 0.015 \\
+ Word-aligned local acoustic fusion & 430k & 1 & 2 & 0.776 $\pm$ 0.0004 & 0.623 $\pm$ 0.0002 & 0.818 $\pm$ 0.0001 & 0.803 $\pm$ 0.0002 & 24.55 $\pm$ 0.012 \\
+ Utterance-level global acoustic fusion & 430k & 1 & 1 & 0.767 $\pm$ 0.0005 & 0.609 $\pm$ 0.0004 & 0.811 $\pm$ 0.0002 & 0.802 $\pm$ 0.0003 & 24.55 $\pm$ 0.014 \\
+ Joint local/global acoustic fusion & 628k & 1 & 2 & \textbf{0.778 $\pm$ 0.0006} & \textbf{0.626 $\pm$ 0.0005} & \textbf{0.819 $\pm$ 0.0002} & \textbf{0.806 $\pm$ 0.0002} & \textbf{24.39 $\pm$ 0.018} \\
\bottomrule
\end{tabular}%
}
\end{table*}

\begin{table*}[t]
\centering
\caption{Severity-wise comparison on the official CPC3 evaluation set (five-seed means).}
\label{tab:severity}
\scriptsize
\begin{tabular}{llcccc}
\toprule
Severity & System & F1 & MCC & \corr & \rmse \\
\midrule
\multirow{4}{*}{Mild}
& Text-conditioned decoder baseline & 0.720 & 0.589 & 0.793 & 23.63 \\
& + Word-aligned local acoustic fusion & 0.741 & 0.617 & 0.804 & 23.11 \\
& + Utterance-level global acoustic fusion & 0.723 & 0.593 & 0.798 & 23.40 \\
& + Joint local/global acoustic fusion & \textbf{0.742} & \textbf{0.620} & \textbf{0.807} & \textbf{22.98} \\
\midrule
\multirow{4}{*}{Moderate}
& Text-conditioned decoder baseline & 0.771 & 0.599 & 0.794 & 25.12 \\
& + Word-aligned local acoustic fusion & 0.785 & 0.619 & 0.800 & 24.83 \\
& + Utterance-level global acoustic fusion & 0.779 & 0.609 & 0.802 & 24.71 \\
& + Joint local/global acoustic fusion & \textbf{0.788} & \textbf{0.622} & \textbf{0.804} & \textbf{24.67} \\
\midrule
\multirow{4}{*}{Moderately severe}
& Text-conditioned decoder baseline & 0.799 & 0.575 & 0.735 & 29.15 \\
& + Word-aligned local acoustic fusion & 0.813 & 0.600 & 0.751 & 28.64 \\
& + Utterance-level global acoustic fusion & 0.805 & 0.583 & 0.750 & 28.59 \\
& + Joint local/global acoustic fusion & \textbf{0.815} & \textbf{0.602} & \textbf{0.759} & \textbf{28.31} \\
\bottomrule
\end{tabular}
\end{table*}

Table~\ref{tab:severity} shows that this trend is stable across hearing-severity groups. The joint model improves RMSE from 23.63 to 22.98 for mild listeners, from 25.12 to 24.67 for moderate listeners, and from 29.15 to 28.31 for moderately severe listeners. The improvement is largest in the moderately severe group in absolute RMSE, which is consistent with the intuition that local evidence and global calibration become more valuable when the signal is more difficult to interpret. At the same time, all systems share the same severity embedding, so these gains are not simply due to listener metadata; they come from how acoustic evidence is attached to the reference-word targets.

\subsection{Reference Conditioning and Alignment Diagnostics}

Table~\ref{tab:diagnostics} evaluates three design choices. First, replacing subword-BPE all-head alignment with character-based dynamic top-10 head selection improves local fusion from F1 0.772 to 0.776 and RMSE 24.66 to 24.55. The oracle-clean alignment upper bound is only slightly stronger, indicating that the estimated degraded-speech alignment captures much of the useful local structure without using clean-audio information. Second, the hypothesis-derived baseline performs substantially worse than the teacher-forced reference-conditioned baseline. This control is important because it shows that the method is not merely post-processing Whisper's ASR output. Teacher forcing preserves one prediction coordinate per canonical word, allowing the model to express uncertainty about listener perception even when a normal decoder would have committed to a single recognized sequence. Third, the same qualitative pattern persists with Whisper medium. The medium joint model is slightly stronger in F1, MCC, and correlation but not in RMSE, so the main gain should be attributed to the formulation and acoustic fusion, not only to backbone scaling.

\begin{table*}[t]
\centering
\caption{Diagnostic comparisons on the official CPC3 evaluation set. Alignment results are shown for local fusion with Whisper small. Hypothesis-derived baselines align raw Whisper transcriptions to reference words. Backbone scaling uses joint local/global fusion.}
\label{tab:diagnostics}
\scriptsize
\resizebox{\textwidth}{!}{%
\begin{tabular}{llcccc}
\toprule
Analysis & System & F1 & MCC & \corr & \rmse \\
\midrule
\multirow{3}{*}{Alignment quality} & Subword-BPE all-head local alignment & 0.772 $\pm$ 0.0008 & 0.616 $\pm$ 0.0008 & 0.800 $\pm$ 0.0002 & 24.66 $\pm$ 0.013 \\
& Character-based dynamic top-10 alignment & 0.776 $\pm$ 0.0004 & 0.623 $\pm$ 0.0002 & 0.803 $\pm$ 0.0002 & 24.55 $\pm$ 0.012 \\
& Oracle-clean alignment & 0.777 $\pm$ 0.0008 & 0.624 $\pm$ 0.0008 & 0.804 $\pm$ 0.0002 & 24.49 $\pm$ 0.011 \\
\midrule
\multirow{2}{*}{Reference conditioning} & Whisper-small hypothesis-derived & 0.723 & 0.553 & 0.707 & 31.32 \\
& Whisper-small teacher-forced baseline & 0.760 & 0.601 & 0.795 & 24.92 \\
\midrule
\multirow{2}{*}{Backbone scaling} & Whisper-small joint fusion & 0.778 $\pm$ 0.0006 & 0.626 $\pm$ 0.0005 & 0.806 $\pm$ 0.0002 & 24.39 $\pm$ 0.018 \\
& Whisper-medium joint fusion & 0.781 $\pm$ 0.0002 & 0.628 $\pm$ 0.0006 & 0.807 $\pm$ 0.0003 & 24.41 $\pm$ 0.019 \\
\bottomrule
\end{tabular}%
}
\end{table*}

\subsection{Why Local and Global Evidence Are Complementary}

The local and global branches play different roles. The local branch is word-specific: for each reference word, it asks whether the encoder frames selected by transcript-conditioned alignment provide acoustic support for that word. This is useful for substitutions and omissions. In a substitution-like case, the teacher-forced decoder state still encodes the intended reference word, but the local encoder summary may carry evidence closer to another lexical item. In an omission-like case, there may be no true realization of the reference word; the normalized attention profile then functions as a soft context selector, and weak or diffuse local evidence becomes an informative mismatch signal.

The global branch cannot identify which individual word failed, but it summarizes utterance-level difficulty such as overall distortion, noise, hearing-aid processing artifacts, and calibration-relevant context. This explains why it improves sentence-level RMSE while producing smaller word-level gains. The joint model benefits from both signals: local features sharpen word correctness discrimination, while global features stabilize the conversion from word probabilities to the final sentence percentage.

This interpretation clarifies the relationship with ASR confidence estimation. Standard confidence estimation predicts whether a generated hypothesis word is correct. Here, the target is the perceptual correctness of a canonical reference word for a hearing-impaired listener. The reference-conditioned decoder supplies the lexical coordinate, and the acoustic branches supply evidence about audibility and signal support. This distinction is central in CPC-style scoring, where the desired output is not an ASR transcript but a listener-dependent intelligibility percentage.

\subsection{Target Construction and Reproducibility}

Several details are important because they affect both the target labels and the acoustic pooling. Transcript normalization determines which listener responses are treated as exact matches, and word-level Levenshtein alignment determines whether substitutions and deletions become incorrect reference-word labels. Insertions are excluded because the target space consists only of canonical reference positions. This choice matches percentage-correct scoring over reference words, but it should not be confused with a full ASR word-error-rate objective.

Token-to-word and character-to-word mappings are equally important. The main decoder features are computed with Whisper's standard subword BPE tokenization, while the auxiliary character-level view is used only for alignment extraction. Keeping these two roles separate avoids changing the semantic decoder representation while still benefiting from sharper character-level attention. For reproducibility, reported values are five-seed means with fixed fold assignments, and checkpoint selection uses internal scene-level grouped cross-validation rather than the official evaluation set.

\subsection{From Word Correctness to Sentence Scores}

Equation~\eqref{eq:sent_score} deliberately uses an unweighted masked mean because CPC-style percentage-correct scoring counts reference words uniformly. This choice makes the sentence prediction directly interpretable: every predicted word probability contributes linearly to the final score. It also explains why word-level and sentence-level metrics do not always move in perfect lockstep. A model can correct many local decisions, improving incorrect-word F1, while the resulting RMSE change depends on utterance length, listener variability, and where those corrected decisions occur in the score distribution.

The global branch can be viewed as calibrating this aggregation. The local branch changes the evidence used to estimate each $p_i$, whereas the global feature gives the classifier access to overall utterance difficulty when mapping local word evidence to probabilities. This design is useful because two words with similar local evidence may have different perceptual reliability in globally easy and globally difficult utterances. Joint fusion therefore improves not only local discrimination, but also the consistency of the final average score.

\subsection{Comparison Scope}

The main evidence for the proposed design is the controlled comparison among decoder-only, local-fusion, global-fusion, and joint-fusion variants. These systems share the same frozen backbone, severity conditioning, cross-validation folds, optimizer, and checkpoint-selection rule; the differences in Table~\ref{tab:main} therefore isolate how acoustic evidence is attached to the reference-word targets. This is more informative for the present paper than a broad leaderboard-style comparison, because text-assisted systems and fully non-intrusive systems operate under different input assumptions. The reported numbers should therefore be read as a controlled study of the text-assisted formulation, not as a claim of superiority over systems that do not use the canonical transcript.

The diagnostic conditions should also be read with this scope in mind. The oracle-clean alignment setting is not a deployable system but a ceiling for the local alignment path. Likewise, the hypothesis-derived baseline is not intended as a strong ASR system; it tests whether simply decoding with Whisper and aligning the transcript can explain the gain. The large gap between that baseline and teacher-forced reference conditioning supports the central claim that preserving canonical word positions is beneficial for this task.

\subsection{Limitations and Future Work}

The main practical limitation is computational cost. The local and joint variants share one encoder pass with the decoder baseline but require an auxiliary decoder-only pass for character-level alignment extraction. The overhead can exceed the pass-count proxy because the character sequence is usually longer than the BPE sequence. Future work should distill the alignment branch into a smaller module, cache transcript-conditioned alignment features for repeated evaluation, or replace the auxiliary character decoder with a lightweight monotonic aligner.

A second limitation is the dependence on a known canonical transcript at inference time. This is appropriate for text-assisted intelligibility prediction, fixed-prompt speech tests, and controlled hearing-aid or enhancement-system evaluation, but it is not a general open-world monitoring system. A hybrid system could first identify the expected prompt or obtain a reliable transcript and then apply reference-conditioned correctness prediction. The current listener representation is also coarse: the severity embedding does not encode full audiograms, cognitive factors, or listener-specific lexical familiarity. Extending the metadata branch is a natural direction for improving calibration while keeping the reference-word formulation unchanged.

A remaining experimental limitation is that the diagnostics do not fully isolate lexical priors from acoustic evidence. The hypothesis-derived baseline tests whether raw Whisper decoding explains the gain, and the local/global ablations show that added acoustic paths are useful. However, stricter no-audio or shuffled-audio controls, together with a same-backbone direct sentence-regression baseline, would more directly quantify the effects of word-level aggregation, canonical transcript priors, and acoustic evidence.

Finally, the model treats all evaluated reference words equally when forming the sentence score. This matches the official percentage-correct target, but it does not distinguish function words, content words, phonetic confusability, or perceptually weak consonantal regions. Future analyses could test whether local acoustic fusion is especially beneficial for short words, low-energy segments, or words adjacent to noise bursts. Such analysis would make the model more useful as a diagnostic tool for enhancement systems, not only as a score predictor.

\section{Conclusion}

We presented a text-assisted intelligibility prediction framework based on reference-conditioned word-level correctness modeling. The method predicts correctness at canonical reference-word positions and derives sentence-level intelligibility by masked averaging. On top of a frozen Whisper teacher-forced decoder baseline, alignment-aware local acoustic fusion improves word-level discrimination, utterance-level global fusion improves calibration, and their combination gives the strongest overall result on the official CPC3 evaluation set. Diagnostic analyses show that character-level dynamic cross-attention alignment is useful, that teacher-forced reference conditioning is substantially stronger than hypothesis-derived post-processing, and that the qualitative trend is preserved across Whisper small and medium. Future work should add direct utterance-level and no-audio controls, reduce alignment-extraction cost, incorporate richer listener metadata, and analyze which lexical and acoustic conditions benefit most from local fusion.

\section*{Acknowledgment}
This manuscript used generative AI for English editing and wording suggestions in the manuscript text. All scientific claims, experiments, and final text were reviewed and validated by the authors, who take responsibility for the submitted manuscript.

\clearpage
\bibliographystyle{IEEEtran}
\bibliography{refs_slt_revised}

\end{document}